\documentclass{mem}
\usepackage{natbib}\usepackage{txfonts}\usepackage{balance}
\usepackage{graphicx}
\usepackage[a4paper]{hyperref}
\idline{75}{1}
\begin{document}

\graphicspath{{./fig/}{./png/}}
%|||||||||||||||||||||||||||||||||||||||||||||||||||||||||||||||||||
%             Customized Commands
%|||||||||||||||||||||||||||||||||||||||||||||||||||||||||||||||||||
%  mathematical abbreviations
%  =========================

% math defs
\newcommand{\EQ}{\begin{equation}}
\newcommand{\EN}{\end{equation}}
\newcommand{\EQA}{\begin{eqnarray}}
\newcommand{\ENA}{\end{eqnarray}}
\newcommand{\eq}[1]{(\ref{#1})}
\newcommand{\EEq}[1]{Equation~(\ref{#1})}
\newcommand{\Eq}[1]{equation~(\ref{#1})}
\newcommand{\Eqs}[2]{equations~(\ref{#1}) and~(\ref{#2})}
\newcommand{\Eqss}[2]{equations~(\ref{#1})--(\ref{#2})}
\newcommand{\eqs}[2]{(\ref{#1}) and~(\ref{#2})}
\newcommand{\App}[1]{Appendix~\ref{#1}}
\newcommand{\Sec}[1]{\S\ref{#1}}
\newcommand{\Secs}[2]{\S\S\ref{#1} and \ref{#2}}
\newcommand{\Fig}[1]{Fig.~\ref{#1}}
\newcommand{\FFig}[1]{Figure~\ref{#1}}
\newcommand{\Tab}[1]{Table~\ref{#1}}
\newcommand{\Figs}[2]{Figs~\ref{#1} and \ref{#2}}
\newcommand{\Figss}[2]{Figs~\ref{#1}--\ref{#2}}
\newcommand{\Tabs}[2]{Tables~\ref{#1} and \ref{#2}}
\newcommand{\bra}[1]{\langle #1\rangle}
\newcommand{\bbra}[1]{\left\langle #1\right\rangle}
\newcommand{\mean}[1]{\overline #1}
\newcommand{\meanemf}{\overline{\mbox{\boldmath ${\cal E}$}}{}}{}
\newcommand{\meanFF}{\overline{\mbox{\boldmath ${\cal F}$}}{}}{}
\newcommand{\meanemfs}{\overline{\cal E} {}}
\newcommand{\meanSSSS}{\overline{\mathsf{S}}}
\newcommand{\meanuu}{\overline{\mbox{\boldmath $u$}}{}}{}
\newcommand{\meanoo}{\overline{\mbox{\boldmath $\omega$}}{}}{}
\newcommand{\meanE}{\overline{{\cal E}}}
\newcommand{\meanEE}{\overline{\mbox{\boldmath ${\cal E}$}}{}}{}
\newcommand{\meanEMF}{\overline{\mbox{\boldmath ${\cal E}$}}{}}{}
\newcommand{\meanuxB}{\overline{\mbox{\boldmath $\delta u\times \delta B$}}{}}{}
\newcommand{\meanJB}{\overline{\mbox{\boldmath $J\cdot B$}}{}}{}
\newcommand{\meanAB}{\overline{\mbox{\boldmath $A\cdot B$}}{}}{}
\newcommand{\meanjb}{\overline{\mbox{\boldmath $j\cdot b$}}{}}{}
\newcommand{\meanBB}{\overline{\mbox{\boldmath $B$}}{}}{}
\newcommand{\meanJJ}{\overline{\mbox{\boldmath $J$}}{}}{}
\newcommand{\meanUU}{\overline{\mbox{\boldmath $U$}}{}}{}
\newcommand{\meanWW}{\overline{\mbox{\boldmath $W$}}{}}{}
\newcommand{\meanA}{\overline{A}}
\newcommand{\meanB}{\overline{B}}
\newcommand{\meanC}{\overline{C}}
\newcommand{\meanU}{\overline{U}}
\newcommand{\meanR}{\overline{\rho}}
\newcommand{\meanJ}{\overline{J}}
\newcommand{\meanS}{\overline{S}}
\newcommand{\meanF}{\overline{\cal F}}
%
% tilde
%
\newcommand{\teps}{\tilde{\epsilon} {}}
\newcommand{\zh}{\hat{z}}
%
%  unit vectors
%
\newcommand{\pphi}{\hat{\bm{\phi}}}
\newcommand{\eee}{\hat{\mbox{\boldmath $e$}} {}}
\newcommand{\nnn}{\hat{\mbox{\boldmath $n$}} {}}
\newcommand{\rrr}{\hat{\mbox{\boldmath $r$}} {}}
\newcommand{\vvv}{\hat{\mbox{\boldmath $v$}} {}}
\newcommand{\xxx}{\hat{\mbox{\boldmath $x$}} {}}
\newcommand{\yyy}{\hat{\mbox{\boldmath $y$}} {}}
\newcommand{\zzz}{\hat{\mbox{\boldmath $z$}} {}}
\newcommand{\ttt}{\hat{\mbox{\boldmath $\theta$}} {}}
\newcommand{\OOO}{\hat{\mbox{\boldmath $\Omega$}} {}}
\newcommand{\ooo}{\hat{\mbox{\boldmath $\omega$}} {}}
\newcommand{\BBBB}{\hat{\mbox{\boldmath $B$}} {}}
\newcommand{\Bhat}{\hat{B}}
%
%  vectors
%
\newcommand{\gggg}{\mbox{\boldmath $g$} {}}
\newcommand{\ddd}{\mbox{\boldmath $d$} {}}
\newcommand{\rr}{\mbox{\boldmath $r$} {}}
\newcommand{\yy}{\mbox{\boldmath $y$} {}}
\newcommand{\zz}{\mbox{\boldmath $z$} {}}
\newcommand{\vv}{\mbox{\boldmath $v$} {}}
\newcommand{\ww}{\mbox{\boldmath $w$} {}}
\newcommand{\mm}{\mbox{\boldmath $m$} {}}
\newcommand{\PP}{\mbox{\boldmath $P$} {}}
\newcommand{\bp}{\mbox{\boldmath $p$} {}}
\newcommand{\pp}{\mbox{\boldmath $p$} {}}
\newcommand{\II}{\mbox{\boldmath $I$} {}}
\newcommand{\RR}{\mbox{\boldmath $R$} {}}
\newcommand{\kk}{\mbox{\boldmath $k$} {}}
\newcommand{\KK}{\mbox{\boldmath $K$} {}}
\newcommand{\uu}{\mbox{\boldmath $u$} {}}
\newcommand{\UU}{\mbox{\boldmath $U$} {}}
\newcommand{\xx}{\mbox{\boldmath $x$} {}}
\newcommand{\bb}{\mbox{\boldmath $b$} {}}
\newcommand{\BB}{\mbox{\boldmath $B$} {}}
\newcommand{\EE}{\mbox{\boldmath $E$} {}}
\newcommand{\jj}{\mbox{\boldmath $j$} {}}
\newcommand{\JJ}{\mbox{\boldmath $J$} {}}
\newcommand{\SSS}{\mbox{\boldmath $S$} {}}
\newcommand{\AAA}{\mbox{\boldmath $A$} {}}
\newcommand{\aaaa}{\mbox{\boldmath $a$} {}}
\newcommand{\ee}{\mbox{\boldmath $e$} {}}
\newcommand{\nn}{\mbox{\boldmath $n$} {}}
\newcommand{\ff}{\mbox{\boldmath $f$} {}}
\newcommand{\hh}{\mbox{\boldmath $h$} {}}
\newcommand{\FF}{\mbox{\boldmath $F$} {}}
\newcommand{\EEE}{\mbox{\boldmath ${\cal E}$} {}}
\newcommand{\FFF}{\mbox{\boldmath ${\cal F}$} {}}
\newcommand{\TT}{\mbox{\boldmath $T$} {}}
\newcommand{\MM}{\mbox{\boldmath $M$} {}}
\newcommand{\GG}{\mbox{\boldmath $G$} {}}
\newcommand{\WW}{\mbox{\boldmath $W$} {}}
\newcommand{\QQ}{\mbox{\boldmath $Q$} {}}
\newcommand{\grav}{\mbox{\boldmath $g$} {}}
\newcommand{\nab}{\mbox{\boldmath $\nabla$} {}}
\newcommand{\OO}{\mbox{\boldmath $\Omega$} {}}
\newcommand{\oo}{\mbox{\boldmath $\omega$} {}}
\newcommand{\ttau}{\mbox{\boldmath $\tau$} {}}
\newcommand{\LL}{\mbox{\boldmath $\Lambda$} {}}
\newcommand{\llambda}{\mbox{\boldmath $\lambda$} {}}
\newcommand{\pomega}{\mbox{\boldmath $\varpi$} {}}
%
%  tensors
%
\newcommand{\DDDD}{\mbox{\boldmath ${\sf D}$} {}}
\newcommand{\IIII}{\mbox{\boldmath ${\sf I}$} {}}
\newcommand{\LLLL}{\mbox{\boldmath ${\sf L}$} {}}
\newcommand{\MMMM}{\mbox{\boldmath ${\sf M}$} {}}
\newcommand{\PPPP}{\mbox{\boldmath ${\sf P}$} {}}
\newcommand{\QQQQ}{\mbox{\boldmath ${\sf Q}$} {}}
\newcommand{\RRRR}{\mbox{\boldmath ${\sf R}$} {}}
\newcommand{\SSSS}{\mbox{\boldmath ${\sf S}$} {}}
\newcommand{\BBBBB}{\mbox{\boldmath ${\sf B}$} {}}
\newcommand{\tAAAA}{\tilde{\mbox{\boldmath ${\sf A}$}} {}}
\newcommand{\tDDDD}{\tilde{\mbox{\boldmath ${\sf D}$}} {}}
\newcommand{\tRRRR}{\tilde{\mbox{\boldmath ${\sf R}$}} {}}
\newcommand{\tQQQQ}{\tilde{\mbox{\boldmath ${\sf Q}$}} {}}
\newcommand{\AAAA}{\mbox{\boldmath ${\cal A}$} {}}
\newcommand{\BBB}{\mbox{\boldmath ${\cal B}$} {}}
\newcommand{\emf}{\mbox{\boldmath ${\cal E}$} {}}
\newcommand{\GGG}{\mbox{\boldmath ${\cal G}$} {}}
\newcommand{\HHH}{\mbox{\boldmath ${\cal H}$} {}}
\newcommand{\QQQ}{\mbox{\boldmath ${\cal Q}$} {}}
\newcommand{\GGGG}{{\bf G} {}}
%
%  operators, subscripts, etc  (roman)
%
\newcommand{\ii}{{\rm i}}
\newcommand{\erf}{{\rm erf}}
\newcommand{\grad}{{\rm grad} \, {}}
\newcommand{\curl}{{\rm curl} \, {}}
\newcommand{\dive}{{\rm div}  \, {}}
\newcommand{\Dive}{{\rm Div}  \, {}}
\newcommand{\sgn}{{\rm sgn}  \, {}}
\newcommand{\DD}{{\rm D} {}}
\newcommand{\DDD}{{\cal D} {}}
\newcommand{\dd}{{\rm d} {}}
\newcommand{\const}{{\rm const}  {}}
\newcommand{\CR}{{\rm CR}}
\def\degr{\hbox{$^\circ$}}
\def\la{\mathrel{\mathchoice {\vcenter{\offinterlineskip\halign{\hfil
$\displaystyle##$\hfil\cr<\cr\sim\cr}}}
{\vcenter{\offinterlineskip\halign{\hfil$\textstyle##$\hfil\cr<\cr\sim\cr}}}
{\vcenter{\offinterlineskip\halign{\hfil$\scriptstyle##$\hfil\cr<\cr\sim\cr}}}
{\vcenter{\offinterlineskip\halign{\hfil$\scriptscriptstyle##$\hfil\cr<\cr\sim\cr}}}}}
\def\ga{\mathrel{\mathchoice {\vcenter{\offinterlineskip\halign{\hfil
$\displaystyle##$\hfil\cr>\cr\sim\cr}}}
{\vcenter{\offinterlineskip\halign{\hfil$\textstyle##$\hfil\cr>\cr\sim\cr}}}
{\vcenter{\offinterlineskip\halign{\hfil$\scriptstyle##$\hfil\cr>\cr\sim\cr}}}
{\vcenter{\offinterlineskip\halign{\hfil$\scriptscriptstyle##$\hfil\cr>\cr\sim\cr}}}}}
%
%  numbers
%
\def\Ta{\mbox{\rm Ta}}
\def\Ra{\mbox{\rm Ra}}
\def\Ma{\mbox{\rm Ma}}
\def\Co{\mbox{\rm Co}}
\def\Roo{\mbox{\rm Ro}^{-1}}
\def\Rooo{\mbox{\rm Ro}^{-2}}
\def\Pra{\mbox{\rm Pr}}
\def\Pran{\mbox{\rm Pr}}
\def\Pm{\mbox{\rm Pr}_M}
\def\Rm{\mbox{\rm Re}_M}
\def\Rey{\mbox{\rm Re}}
\def\Pe{\mbox{\rm Pe}}
\newcommand{\ea}{{\rm et al.\ }}
\newcommand{\eaa}{{\rm et al.\ }}
\def\half{{\textstyle{1\over2}}}
\def\threehalf{{\textstyle{3\over2}}}
\def\threequarter{{\textstyle{3\over4}}}
\def\sevenhalf{{\textstyle{7\over2}}}
\def\onethird{{\textstyle{1\over3}}}
\def\onesixth{{\textstyle{1\over6}}}
\def\twothird{{\textstyle{2\over3}}}
\def\fourthird{{\textstyle{4\over3}}}
\def\quarter{{\textstyle{1\over4}}}
\newcommand{\W}{\,{\rm W}}
\newcommand{\V}{\,{\rm V}}
\newcommand{\kV}{\,{\rm kV}}
\newcommand{\T}{\,{\rm T}}
\newcommand{\G}{\,{\rm G}}
\newcommand{\Hz}{\,{\rm Hz}}
\newcommand{\nHz}{\,{\rm nHz}}
\newcommand{\kHz}{\,{\rm kHz}}
\newcommand{\kG}{\,{\rm kG}}
\newcommand{\K}{\,{\rm K}}
\newcommand{\g}{\,{\rm g}}
\newcommand{\s}{\,{\rm s}}
\newcommand{\mpers}{\,{\rm m/s}}
\newcommand{\ks}{\,{\rm ks}}
\newcommand{\cm}{\,{\rm cm}}
\newcommand{\cmcube}{\,{\rm cm^{-3}}} 
\newcommand{\m}{\,{\rm m}}
\newcommand{\km}{\,{\rm km}}
\newcommand{\cms}{\,{\rm cm/s}}
\newcommand{\kms}{\,{\rm km/s}}
\newcommand{\gpercc}{\,{\rm g/cm}^3}
\newcommand{\kg}{\,{\rm kg}}
\newcommand{\ug}{\,\mu{\rm g}}
\newcommand{\kW}{\,{\rm kW}}
\newcommand{\MW}{\,{\rm MW}}
\newcommand{\Mm}{\,{\rm Mm}}
\newcommand{\Mx}{\,{\rm Mx}}
\newcommand{\pc}{\,{\rm pc}}
\newcommand{\kpc}{\,{\rm kpc}}
\newcommand{\yr}{\,{\rm yr}}
\newcommand{\Myr}{\,{\rm Myr}}
\newcommand{\Gyr}{\,{\rm Gyr}}
\newcommand{\erg}{\,{\rm erg}}
\newcommand{\mol}{\,{\rm mol}}
\newcommand{\dyn}{\,{\rm dyn}}
\newcommand{\J}{\,{\rm J}}
\newcommand{\RM}{\,{\rm RM}}
\newcommand{\EM}{\,{\rm EM}}
\newcommand{\AU}{\,{\rm AU}}
\newcommand{\A}{\,{\rm A}}
%\newcommand{\kA}{\,{\rm kA}}
%%%%%%%%%%%%%%%%%%%%%%%%%%%%%%%%%%%%%%%%%%%%%%%%%%%%%%%%%%%%%%%%%%%%%%%%
%
%  journals
%
\newcommand{\yastroph}[2]{ #1, astro-ph/#2}
\newcommand{\ycsf}[3]{ #1, {Chaos, Solitons \& Fractals,} {#2}, #3}
\newcommand{\yepl}[3]{ #1, {Europhys.\ Lett.,} {#2}, #3}
\newcommand{\yaj}[3]{ #1, {AJ,} {#2}, #3}
\newcommand{\yjgr}[3]{ #1, {J.\ Geophys.\ Res.,} {#2}, #3}
\newcommand{\ysol}[3]{ #1, {Sol.\ Phys.,} {#2}, #3}
\newcommand{\yapj}[3]{ #1, {ApJ,} {#2}, #3}
\newcommand{\ypasp}[3]{ #1, {PASP,} {#2}, #3}
\newcommand{\yapjl}[3]{ #1, {ApJ,} {#2}, #3}
\newcommand{\yapjs}[3]{ #1, {ApJS,} {#2}, #3}
\newcommand{\yan}[3]{ #1, {Astron.\ Nachr.,} {#2}, #3}
\newcommand{\yzfa}[3]{ #1, {Z.\ f.\ Ap.,} {#2}, #3}
\newcommand{\ymhdn}[3]{ #1, {Magnetohydrodyn.} {#2}, #3}
\newcommand{\yana}[3]{ #1, {A\&A,} {#2}, #3}
\newcommand{\yanas}[3]{ #1, {A\&AS,} {#2}, #3}
\newcommand{\yanar}[3]{ #1, {A\&A Rev.,} {#2}, #3}
\newcommand{\yass}[3]{ #1, {Ap\&SS,} {#2}, #3}
\newcommand{\ygafd}[3]{ #1, {Geophys.\ Astrophys.\ Fluid Dyn.,} {#2}, #3}
\newcommand{\ypasj}[3]{ #1, {Publ.\ Astron.\ Soc.\ Japan,} {#2}, #3}
\newcommand{\yjfm}[3]{ #1, {J.\ Fluid Mech.,} {#2}, #3}
\newcommand{\ypepi}[3]{ #1, {Phys.\ Earth Planet.\ Int.,} {#2}, #3}
\newcommand{\ypf}[3]{ #1, {Phys.\ Fluids,} {#2}, #3}
\newcommand{\ypp}[3]{ #1, {Phys.\ Plasmas,} {#2}, #3}
\newcommand{\ysov}[3]{ #1, {Sov.\ Astron.,} {#2}, #3}
\newcommand{\ysovl}[3]{ #1, {Sov.\ Astron.\ Lett.,} {#2}, #3}
\newcommand{\yjetp}[3]{ #1, {Sov.\ Phys.\ JETP,} {#2}, #3}
\newcommand{\yphy}[3]{ #1, {Physica,} {#2}, #3}
\newcommand{\yaraa}[3]{ #1, {ARA\&A,} {#2}, #3}
\newcommand{\yprs}[3]{ #1, {Proc.\ Roy.\ Soc.\ Lond.,} {#2}, #3}
\newcommand{\yprt}[3]{ #1, {Phys.\ Rep.,} {#2}, #3}
\newcommand{\yprl}[3]{ #1, {Phys.\ Rev.\ Lett.,} {#2}, #3}
\newcommand{\yphl}[3]{ #1, {Phys.\ Lett.,} {#2}, #3}
\newcommand{\yptrs}[3]{ #1, {Phil.\ Trans.\ Roy.\ Soc.,} {#2}, #3}
\newcommand{\ymn}[3]{ #1, {MNRAS,} {#2}, #3}
\newcommand{\ynat}[3]{ #1, {Nature,} {#2}, #3}
\newcommand{\ysci}[3]{ #1, {Science,} {#2}, #3}
\newcommand{\ysph}[3]{ #1, {Solar Phys.,} {#2}, #3}
\newcommand{\ypr}[3]{ #1, {Phys.\ Rev.,} {#2}, #3}
\newcommand{\ypre}[3]{ #1, {Phys.\ Rev.\ E,} {#2}, #3}
\newcommand{\ypnas}[3]{ #1, {Proc.\ Nat.\ Acad.\ Sci.,} {#2}, #3}
\newcommand{\yicarus}[3]{ #1, {Icarus,} {#2}, #3}
\newcommand{\yspd}[3]{ #1, {Sov.\ Phys.\ Dokl.,} {#2}, #3}
\newcommand{\yjcp}[3]{ #1, {J.\ Comput.\ Phys.,} {#2}, #3}
\newcommand{\yjour}[4]{ #1, {#2}, {#3}, #4}
\newcommand{\yprep}[2]{ #1, {\sf #2}}
\newcommand{\ybook}[3]{ #1, {#2} (#3)}
\newcommand{\yproc}[5]{ #1, in {#3}, ed.\ #4 (#5), #2}
\newcommand{\pproc}[4]{ #1, in {#2}, ed.\ #3 (#4), (in press)}
\newcommand{\pprocc}[5]{ #1, in {#2}, ed.\ #3 (#4, #5)}
%%%%%%%%%%%%%%%%%%%%%%%%%%%%%%%%%%%%%%%%%%%%%%%%%%%%%%%%%%%%%%%%%%%%%%%%

\title{
Dynamos in accretion discs
}

   \subtitle{}

\author{
A.\ Brandenburg\inst{1} 
\and B.\ von Rekowski\inst{2}
          }

  \offprints{A. Brandenburg}

\institute{
NORDITA, AlbaNova University Center, SE-10691 Stockholm, Sweden
\and
School of Mathematics and Statistics, University of St.\ Andrews,
  North Haugh, St.\ Andrews, Fife KY16 9SS, UK
\email{brandenb@nordita.dk}
}

\authorrunning{Brandenburg \& von Rekowski}

\titlerunning{Dynamos in accretion discs}

\abstract{
It is argued that accretion discs in young stellar objects
may have hot coronae that are heated by magnetic reconnection.
This is a consequence of the magneto-rotational instability
driving turbulence in the disc.
Magnetic reconnection away from the midplane leads to heating
of the corona which, in turn, contributes to driving disc winds.
}

\maketitle{}

\section{Introduction}

Accretion discs provide the channel through which matter from the outside
is able to collapse onto some common central object such as a protostar.
Prior to the collapse, most of the matter will be in random motion
and will have excess angular momentum with respect to the forming star.
The disc allows the matter to get rid of this excess angular momentum,
but in order to do this there needs to be some enhanced viscosity of
some type.
It is now generally accepted that this is provided by turbulence which
is driven by the magneto-rotational instability
(MRI; see Balbus \& Hawley 1991, 1998).

\begin{figure}[t!]
\resizebox{\hsize}{!}{\includegraphics[clip=true]{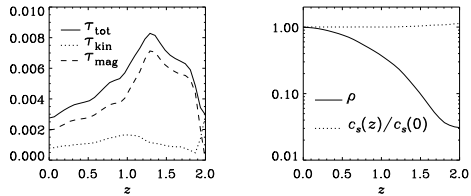}}
\caption{\footnotesize
Dependence of the stress component $\overline{\tau}_{\varpi\phi}$
(here denoted by $\tau_{\rm tot}$), separately for the kinetic and magnetic
contributions, $\tau_{\rm kin}$ and $\tau_{\rm mag}$, respectively,
together with the sum of the two denoted by total (left)
as well as the vertical dependence of density and sound speed (right).
Note that $\tau_{\rm tot}$ is neither proportional to the density $\rho$ nor
the sound speed $c_{\rm s}$.
[Adapted from Brandenburg et al.\ (1996a).]
}
\label{palpz}
\end{figure}

An automatic by-product of MRI-driven turbulence is the possibility of
significant resistive Joule heating in the layers aways from the
midplane, where the density is small and hence the heating per unit
mass is large.
This is demonstrated in \Fig{palpz}, where we show 
that the total (magnetic and kinetic) stress, that leads to
viscous and resistive heating,
does not decrease with height away from the midplane,
even though the product of rate of strain and density does
decrease because of decreasing density.
This was originally demonstrated only for nearly isothermal
discs (Brandenburg et al.\ 1996a), see \Fig{palpz}, but this has
now also been shown for discs with full radiation transfer included
(Turner 2004).

An important expectation from this is that the discs should be surrounded
by hot coronae.
In other words, not only the central protostar, but also the disc itself
should be surrounded by hot coronae that are heated by the associated
Joule (or resistive) heating via magnetic reconnection, as has been
demonstrated in numerical experiments (Galsgaard \& Nordlund 1996).
This leads to the possibility of hot coronae with temperatures close to the
Virial temperature, $T_{\rm vir}$, given by
\EQ
c_pT_{\rm vir}(r)=GM_*/r,
\EN
where $c_p$ is the specific heat at constant pressure, $G$ is Newton's
constant, $M_*$ is the mass of the protostar, and $r$ is the distance
from the central object.
The only process that can possibly counteract the tendency for strong
coronal heating is radiative cooling.
At present it is unclear what temperatures will be achieved if realistic
cooling is included.

\section{The mean accretion stress}

\begin{figure}[t!]
\resizebox{\hsize}{!}{\includegraphics[clip=true]{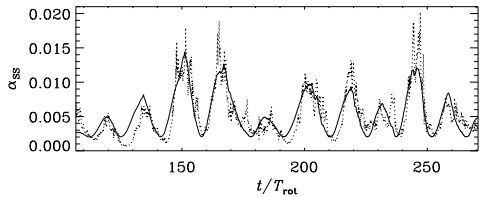}}
\caption{\footnotesize
Time series of the Shakura-Sunyaev viscosity alpha, $\alpha_{\rm SS}$.
Here, time is normalized in terms of the orbital time,
$T_{\rm rot}=2\pi/\Omega$, where $\Omega$ is the local
angular velocity.
Note that $\alpha_{\rm SS}$ fluctuates strongly in time
about an average value of around 0.01.
[Adapted from Brandenburg (1998).]
}
\label{alpt}
\end{figure}

The basic dynamics of the MRI can be modeled using simulations in a
local cartesian geometry using shearing sheet boundary conditions in the radial
direction.
One of the main output parameters is the Shakura--Sunyaev viscosity
parameter, $\alpha_{\rm SS}$, which is a non-dimensional measure
of the turbulent viscosity $\nu_{\rm t}$, in terms of the
sound speed $c_{\rm s}$ and the disc scale height $H$, i.e.\
\EQ
\nu_{\rm t}=\alpha_{\rm SS}c_{\rm s}H.
\EN
Assuming that there is no external torque, this $\alpha_{\rm SS}$
is given by
\EQ
\overline{\ttau}_{\varpi\phi}=
-\rho\nu_{\rm t}\varpi{\partial\overline{\Omega}\over\partial\varpi},
\label{stressMF}
\EN
where $\varpi$ is the cylindrical radius, $\rho$ is the density,
$\overline{\Omega}$ is the mean angular velocity, and
$\overline{\tau}_{\varpi\phi}$ is the total `horizontal' stress given by
\EQ
\overline{\tau}_{\varpi\phi}=-\overline{b_\varpi b_\phi}/\mu_0
+\overline{\rho u_\varpi u_\phi},
\label{stress}
\EN
where $\bb$ and $\uu$ denote the fluctuating components of the
magnetic and velocity fields, and $\mu_0$ is the vacuum permeability.
The first term, i.e.\ the magnetic stress, is usually the largest term.
The resulting stress determined from the simulations and expressed in
nondimensional form as $\alpha_{\rm SS}(t)$, is shown in \Fig{alpt}.
The simulation domain covers the upper disc plane only.

The expression in \Eq{stressMF} would be used in the evolution equation
for the mean angular velocity, $\overline{\Omega}$, where the term
$\nu_{\rm t}$ would simply add to the microscopic viscosity $\nu$
to give a total viscosity, $\nu_{\rm T}=\nu_{\rm t}+\nu$.
On the other hand, \Eq{stress} would be used to determine
$\overline{\tau}_{\varpi\phi}$, and hence $\alpha_{\rm SS}$, from
three-dimensional simulation data.

The value of $\alpha_{\rm SS}(t)$ is notoriously small (around 0.01;
see Hawley et al.\ 1996, Stone et al.\ 1996, Brandenburg et al.\ 1996b).
Observational evidence suggests a typical range
$\alpha_{\rm SS}\approx0.1...0.4$ (King et al.\ 2007).
Larger value of $\alpha_{\rm SS}$ can be obtained if there is an
externally imposed magnetic field.
Also global simulations tend to yield larger values (Hawley 2000).

The variation of $\alpha_{\rm SS}(t)$ shows a typical time scale of
around 15 orbits.
This is here a consequence of the emergence of cyclic large scale
dynamo action.
The magnetic field reverses between two maxima, so the actual period
is actually around 30 orbits.
Physically, this time is related to the turbulent diffusion time.
This is now reasonably well understood and the details of the dynamo
depend sensitively on the boundary conditions.
Such cycles are not expected in global geometry
(Brandenburg \& Campbell 1997).

\section{The mean electromotive force}

An important goal is to simulate outflows from discs that
can produce a magnetic field.
In other words, the magnetic field necessary for launching outflows
is now no longer assumed as being given, but it is self-consistently
generated in the disc, albeit in parameterized form.
This feature was investigated by von Rekowski et al.\ (2003), who studied the
possibility of collimated outflows in the {\em absence} of an ambient magnetic
field.
This idea seems now quite reasonable also from an observational point of view.
M\'enard \& Duch\^ene (2004) report observations of the Taurus-Auriga
molecular cloud which is a nearby star-forming region of low-mass stars.
They find that jets and outflows from
classical T Tauri star--disc systems
in this cloud are not always aligned with the local magnetic field
of the cloud where they are born. This suggests that jets are not likely
to be driven by the cloud's own field.
However, theoretical models have shown that magnetic fields are important
for jet acceleration and collimation.
Furthermore, there is now also observational evidence that jets and outflows
are essentially hydromagnetic in nature. Observations of CO outflows exclude
purely
radiative and thermal driving of jets and outflows
in regions of star formation of low-mass stars (Pudritz 2004).
Thermal pressure might still be dynamically important to lift up the winds
to the critical point.
Moreover, there are observational indications of warm wind regions
with temperatures of a few times $10^4\K$ (G\'omez de Castro \& Verdugo 2003).
These winds cannot be due to stellar radiation because the star is too cool,
suggesting that the warm winds are driven magnetohydrodynamically.
So it is quite plausible that a disc dynamo might be
necessary for the launching of jets and outflows.

In order to address problems such as the launching of disc winds from hot
disc coronae, it is preferable to solve the equations in global geometry.
This increases the computational demands, but some progress can
be made by restricting oneself to axisymmetry.
On the other hand, dynamo action is impossible in axisymmetry
(Cowling 1933).
However, it is sufficient to capture the essential three-dimensional
effects in a statistical sense by considering the averaged equations.

\begin{figure}[t!]
\resizebox{\hsize}{!}{\includegraphics[clip=true]{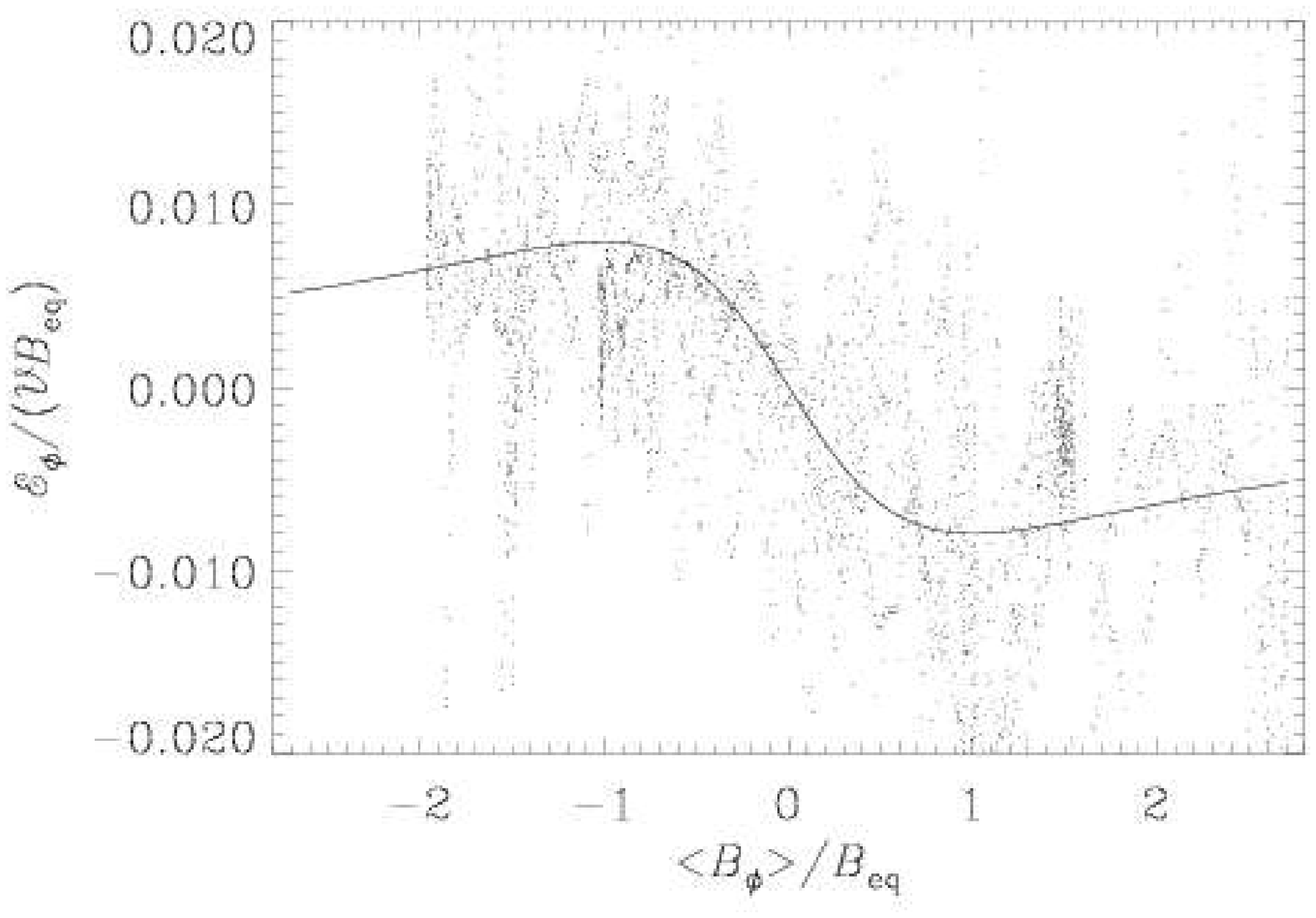}}
\caption{\footnotesize
Scatter plot of the volume averaged toroidal electromotive force,
$\bra{{\cal E}_\phi}$, versus the volume averaged
toroidal magnetic field strength $\bra{B_\phi}$.
Here, $\bra{{\cal E}_\phi}$ is normalized with respect to the rms velocity
${\cal V}$ and the equipartition field strengths
$B_{\rm eq}=\sqrt{\mu_0\rho\uu^2}$.
Both $\bra{{\cal E}_\phi}$ and $\bra{B_\phi}$ are
obtained at different times during the simulation.
Here $\mu_0$ is the vacuum permeability.
Note the negative slope in the diagram around $\bra{B_\phi}=0$,
indicating that the
$\alpha$ effect in mean field dynamo theory is negative.
The solid line represents a fit to a standard quenching law of the form
$-\beta/(1+\beta^2)$, where $\beta=\bra{B_\phi}/B_{\rm eq}$.
[Adapted from Brandenburg \& Donner (1997).]
}
\label{alpb}
\end{figure}

In the averaged equations the nonlinear terms lead to correlations,
just like the stress terms considered in \Eq{stress}.
In the present case there is the mean electromotive force,
\EQ
\meanEMF=\overline{\uu\times\bb},
\label{emf}
\EN
where lower case $\uu$ and $\bb$ denote fluctuations, i.e.\
velocity and magnetic fields are split into mean and fluctuating
components via $\UU=\meanUU+\uu$ and $\BB=\meanBB+\bb$.
\EEq{emf} is used to determine $\meanEMF$ from simulation data.
This expression is thus analogous to \Eq{stress}.

In isotropic turbulence $\meanEMF=0$ (just like there is no off-diagonal
terms in the stress considered above).
However, in the presence of rotation and stratification $\meanEMF$ can be
finite (Krause \& R\"adler 1980).
In particular, theory suggests that $\meanEMF$ has a component along
the direction of the mean field, i.e.\
\EQ
\meanEMF=\alpha_{\rm dyn}\meanBB+...,
\label{emfMF}
\EN
where $\alpha_{\rm dyn}$ is the dynamo alpha which has the dimensions
of a speed.
The dots indicate the presence of additional terms such as turbulent
resistivity, for example.
The expression \Eq{emfMF} would be used in a mean field model
in the averaged induction equation for $\meanBB$, and is thus analogous
to \Eq{stressMF} which is used in the averaged momentum equation.

In \Fig{alpb} we show a scatter plot of the toroidal electromotive force,
$\bra{{\cal E}_\phi}$, versus the volume averaged toroidal magnetic
field strength $\bra{B_\phi}$ that has been obtained from a simulation
of MRI-driven turbulence in the upper disc plane.
Note the negative slope in the diagram indicating that the
$\alpha$ effect in mean field dynamo theory is negative
(Brandenburg et al.\ 1995, Ziegler \& R\"udiger 2000).
This is an unexpected result, because theory predicts $\alpha_{\rm dyn}$ to be
positive in the upper disc plane.
Various proposals for this negative sign have been put forward
(Brandenburg 1998, R\"udiger \& Pipin 2000, Blackman \& Tan 2004).
More detailed analysis confirms the negative value of $\alpha_{\rm dyn}$
and is also able to give its dependence on the distance from the midplane
and other aspects (Brandenburg \& Sokoloff 2002).

There is a great deal of work concerned with the possibility of so-called
catastrophic quenching of $\alpha_{\rm dyn}$ that depends on the value
of the microscopic magnetic Reynolds number.
We cannot go into these aspects here, but we mention only that such
quenching can be alleviated by magnetic helicity fluxes out of the domain
(Blackman \& Field 2000, Kleeorin et al.\ 2000), and that such fluxes
can be driven by shear (Vishniac \& Cho 2001,
Subramanian \& Brandenburg 2004, 2006).
The end result might be an $\alpha$ effect that is no longer
catastrophically quenched; see Brandenburg \& Subramanian (2005)
for a review.
In the following we will just use this as an established fact, as has been
done elsewhere in the literature (Campbell 2000, Bardou et al.\ 2001).

\section{Structured outflows from discs}

\begin{figure}
  \centering
  \includegraphics[height=3.0cm]{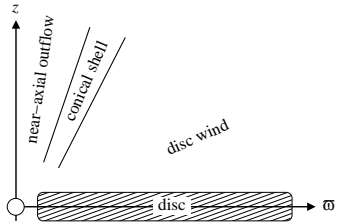}
  \caption{General structure of the outflows typically obtained in
the model of von Rekowski et al.\ (2003), where the cool, dense disc emits
a thermally driven wind and a magneto-centrifugally driven outflow in a
conical shell.
[Adapted from von Rekowski et al.\ (2003).]
  }
  \label{Fig-structure}
\end{figure}

A mean field dynamo model with piece-wise polytropic hydrodynamics
was proposed by von Rekowski et al.\ (2003).
Like Ouyed \& Pudritz (1997a,b) they start with an equilibrium
corona, where they assume constant entropy and hydrostatic equilibrium
according to $c_p T(r)=GM_*/r$.
In order to make the disc cooler, a geometrical
region for the disc is prescribed (see \Fig{Fig-structure}).
An entropy contrast between disc and corona is chosen such that the initial
disc temperature is about $3\times10^3 \K$ in the bulk of the disc.
The low disc temperature corresponds to a high disc density of about
$10^{-10} \dots 10^{-9} \g \cmcube$.
For the disc dynamo, the most important parameter is the dynamo
coefficient $\alpha_{\rm dyn}$ in the mean field induction equation.
The $\alpha$ effect is antisymmetric about the midplane and restricted
to the disc.

\begin{figure}
 \centering
 \includegraphics[height=5.5cm]{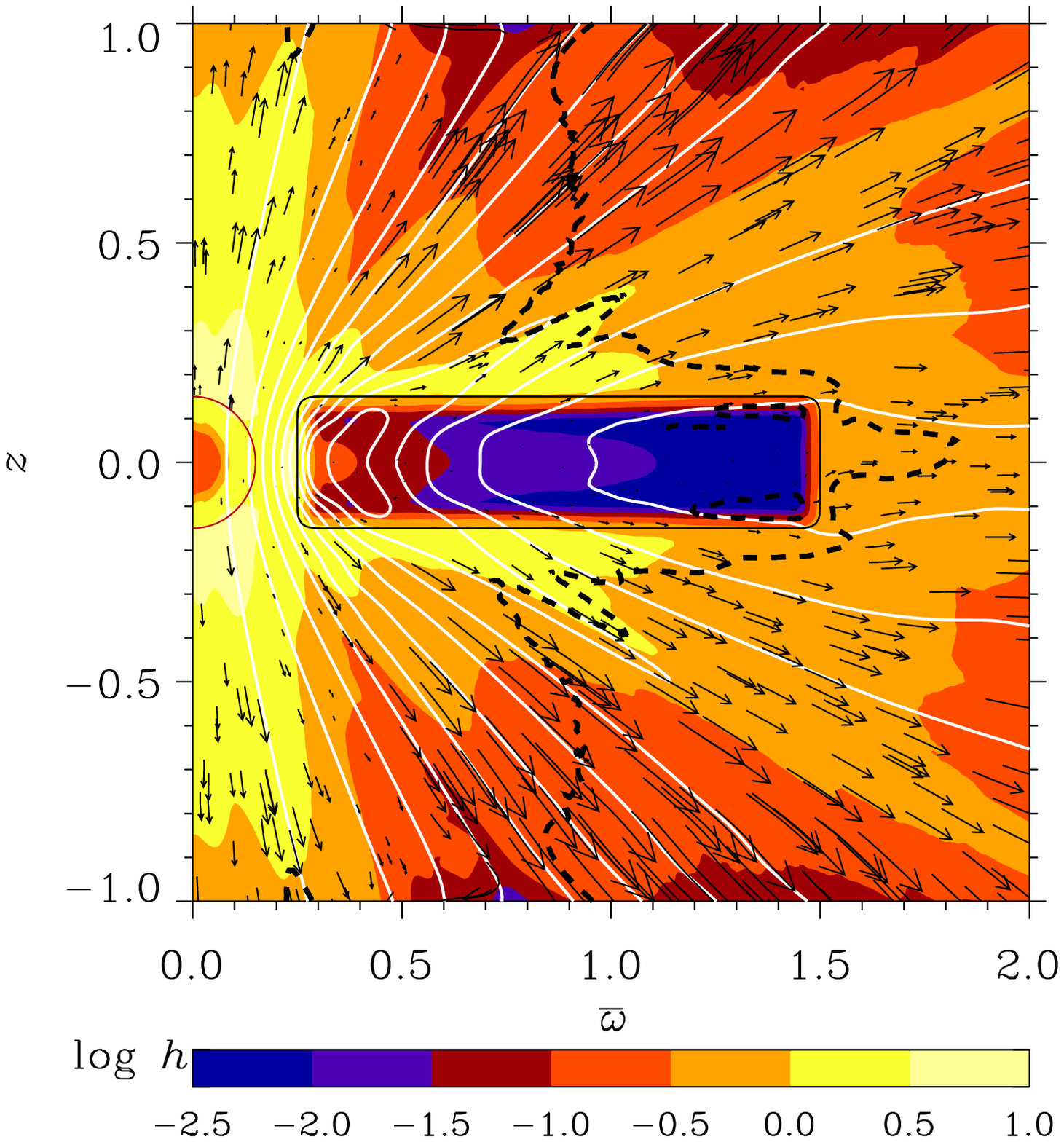}
 \includegraphics[height=5.5cm]{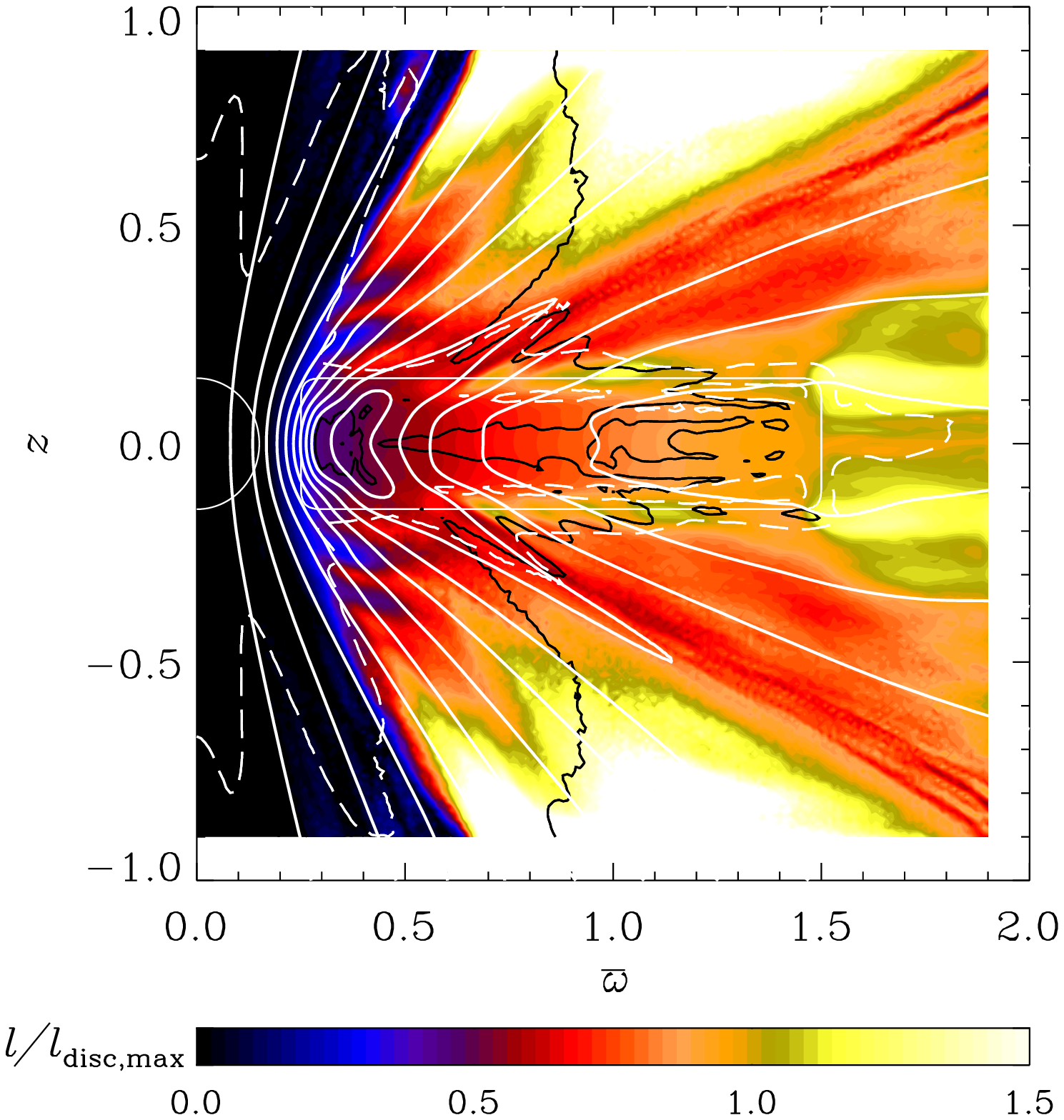}
 \caption{
  Upper panel:
  poloidal velocity vectors and poloidal magnetic field lines (white)
  superimposed on a color scale representation of $\log h$. Specific
  enthalpy $h$ is directly proportional to temperature $T$, and
  $\log h=(-2,-1,0,1)$ corresponds to $T\approx
  (3{\times}10^{3},3{\times}10^{4},3{\times}10^{5},3{\times}10^{6})\,\mbox{K}$.
  The black dashed line shows the fast magnetosonic surface.
  The disc boundary is shown
  with a thin black line, the stellar surface is marked in red.
  The dynamo $\alpha_{\rm dyn}$ coefficient is negative in the upper disc half,
  resulting in roughly dipolar magnetic
  symmetry. Averaged over times $t\approx 897 \dots 906\,{\rm days}$.
  Lower panel:
  color scale representation of the specific
  angular momentum, normalized by the maximum angular momentum in the disc,
  with poloidal magnetic field lines superimposed (white).
  The black solid line shows the Alfv\'en surface, the white dashed
  line the sonic surface.
  Same model as in the upper panel and averaged over same times.
  [Adapted from von Rekowski \& Brandenburg (2004).]
 }
 \label{Fig1}
\end{figure}

The upper panel of Fig.~\ref{Fig1} illustrates that an outflow develops
that has a well-pronounced structure. Within a conical shell originating
from the inner edge of the disc, the terminal outflow speed exceeds
$500\kms$, and temperature and density are lower than elsewhere.
The inner cone around the axis is the hottest and densest region
where the stellar wind speed reaches about $150\kms$. The wind that develops
from the outer parts of the disc has intermediate values of the speed.

The structured outflow is driven by a combination of different processes.
A significant amount of angular momentum is transported outwards
from the disc into the wind along the magnetic field, especially along the
strong lines within the conical shell (see the lower panel of \Fig{Fig1}).
The magnetic field geometry is such that the angle between the rotation
axis and the field lines threading the disc exceeds $30^\circ$ at the disc surface,
which is favorable for magneto-centrifugal acceleration (Blandford \& Payne
1982). However, the Alfv\'en surface is so close to the disc surface
at the outer parts of the disc that acceleration there is mainly due to
the gas-pressure gradient. In the conical shell, however, the outflow is highly
supersonic and yet sub-Alfv\'enic, with the Alfv\'en radius a few times larger
than the radius at the footpoint of the field lines at the disc surface.
The lever arm of about 3 is sufficient for magneto-centrifugal acceleration
to dominate in the conical shell (cf.\ Krasnopolsky et al.\ 1999).

%The inner cone has very low angular momentum and the magnetic lines are
%inclined to the axis by less than $30^\circ$ so that the stellar wind
%can only be gas-pressure driven.
%This picture is confirmed by looking at the ratio of
%magneto-centrifugal to gas-pressure forces, which is much larger than unity
%in the conical shell, where also the poloidal field lines are strongest.
%This is also the region where magnetic pressure
%produced by the toroidal field exceeds gas pressure,
%indicating that the magneto-centrifugally accelerated outflow
%in the conical shell is confined by the toroidal field.

\section{Star--disc coupling}

\begin{figure}[t!]
\resizebox{\hsize}{!}{\includegraphics[clip=true]{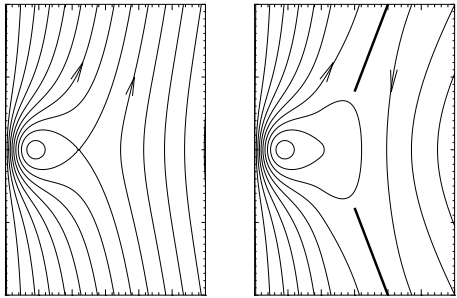}}
\caption{\footnotesize
Sketch showing the formation of an {\sf X}-point when the disc field is
aligned with the dipole (on the left) and the formation of current sheets
with no {\sf X}-point if they are anti-aligned (on the right).
The two current sheets are shown as thick lines.
In the present paper, the second of the two configurations emerges
in all our models,
i.e.\ with current sheets and no {\sf X}-point.
[Adapted from von Rekowski \& Brandenburg (2004).]
}
\label{xpoint}
\end{figure}

The interaction of a stellar magnetic field with a circumstellar accretion
disc and its magnetic field was originally studied in connection with
accretion discs around neutron stars (Ghosh \& Lamb 1979),
but it was later also applied to protostellar magnetospheres
(K\"onigl 1991, Cameron \& Campbell 1993, Shu et al.\ 1994).
Most of the work is based on the assumption that the field in the
disc is constantly being dragged into the inner parts of the
disc from large radii.
The idea behind this is that a magnetized molecular cloud collapses,
in which case the field in the central star and that in the disc are aligned
(Shu et al.\ 1994).
This was studied numerically by Hirose et al.\ (1997) and Miller \& Stone (1997).
In the configurations they considered, there
is an {\sf X}-point in the equatorial plane
(see left hand panel of Fig.~\ref{xpoint}),
which can lead to a strong funnel flow.

The other alternative has been explored by Lovelace et al.\ (1995)
where the magnetic field of the star has been flipped and is now
anti-parallel with the field in the disc, so that the field in the
equatorial plane points in the same direction and has no {\sf X}-point.
However, current sheets develop above and below the disc plane
(see right hand panel of Fig.~\ref{xpoint}).
This configuration is also referred to as the {\sf X}-wind model.
Ironically, this is the field configuration without an {\sf X}-point.

Numerical simulations of such a field configuration
by Hayashi et al.\ (1996) confirm the idea by Lovelace et al.\ (1995)
that closed magnetic loops connecting the star and the disc are twisted
by differential rotation between the star and the disc, and then inflate
to form open stellar and disc field lines (see also Bardou 1999,
Agapitou \& Papaloizou 2000, Uzdensky et al.\ 2002).
Goodson et al.\ (1997,1999) and Goodson \& Winglee (1999) find that
for sufficiently low resistivity, an accretion process develops that is
unsteady and proceeds in an oscillatory fashion.
The same result has also been obtained by von Rekowski \& Brandenburg (2004).
It should be emphasized that in their case the resulting field geometry
is always the second one in Fig.~\ref{xpoint}, i.e.\ the one with current
sheets and no {\sf X}-point.
The inflating magnetosphere expands to
larger radii where matter can be loaded onto the field lines and be ejected as
stellar and disc winds. Reconnection of magnetic field lines allows matter
to flow along them and accrete onto the protostar, in the form of a funnel flow
(see also Romanova et al.\ 2002). Consequently, the outflows show
episodic behavior; see also Matt et al.\ (2002) and
von Rekowski \& Brandenburg (2004).

\begin{figure*}[t!]
\resizebox{.48\hsize}{!}{\includegraphics[clip=true]{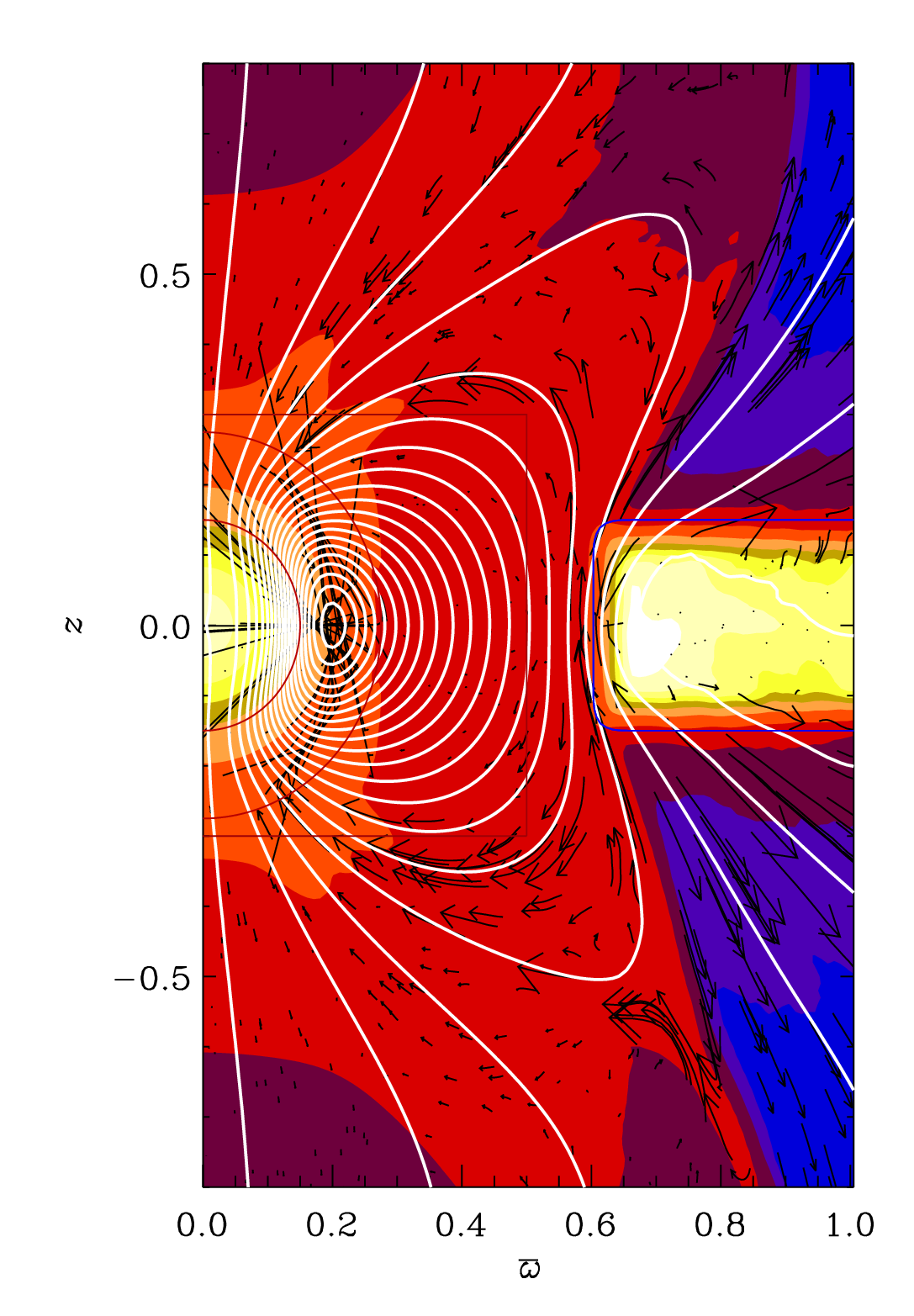}}
\resizebox{.48\hsize}{!}{\includegraphics[clip=true]{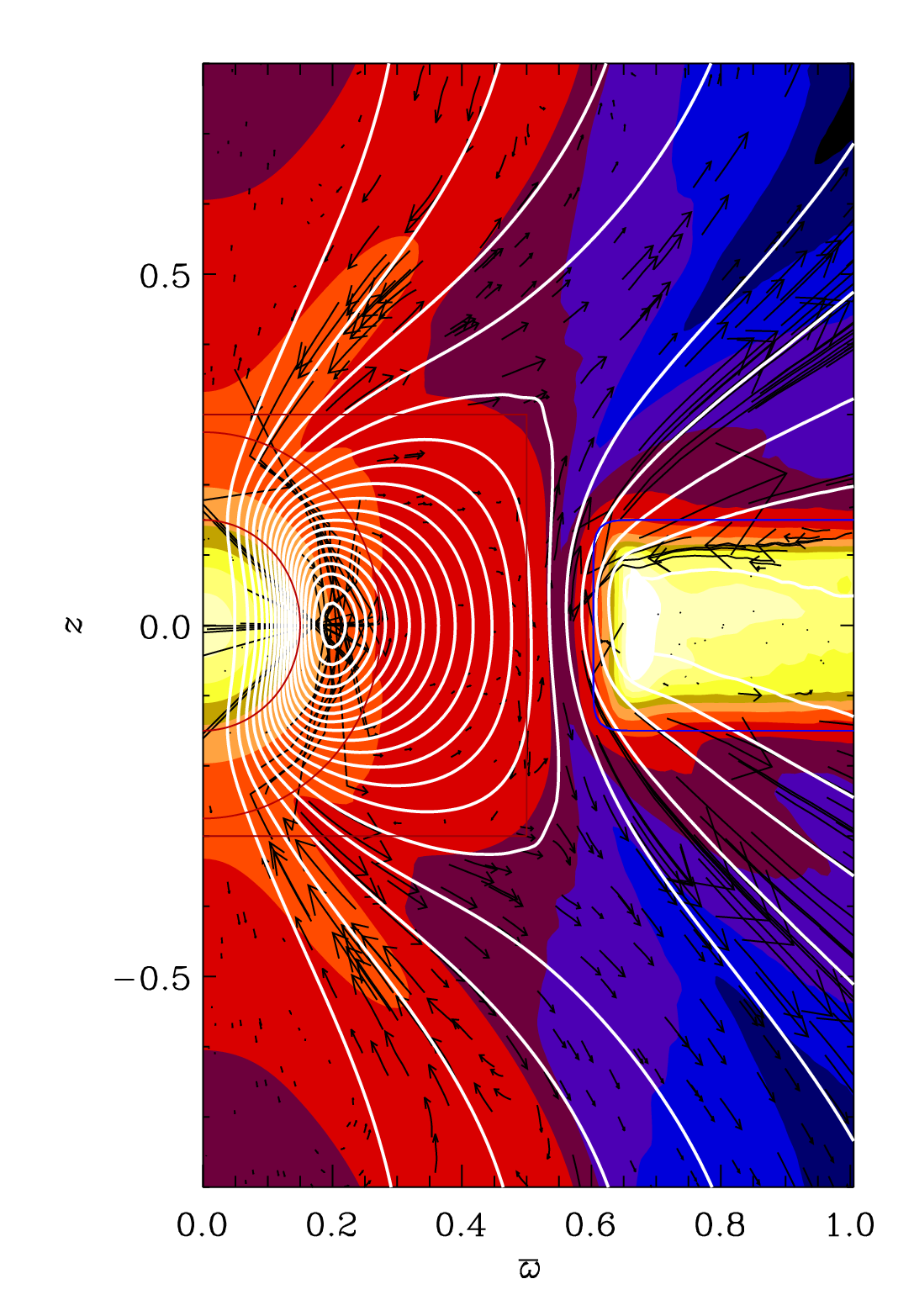}}
\caption{\footnotesize
Colour/grey scale representation of the density (bright colors or light shades
indicate high values; dark colors or dark shades indicate low values)
with poloidal magnetic field lines superimposed (white) and
the azimuthally integrated mass flux density, represented as the vector
$2\pi\varpi\varrho(u_\varpi,u_z)$, shown with arrows
(except in the disc where the density is high and the mass flux vectors
would be too long).
The left hand panel is at the time $t=98$ when the star is magnetically
connected to the disc and the right hand panel is at the time $t=102$ when
the star is magnetically disconnected from the disc.
[Adapted from von Rekowski \& Brandenburg (2004).]
}
\label{FRun_mag_strong-p-98}
\end{figure*}

In the model of von Rekowski \& Brandenburg (2004)
the highly episodic accretion flow and accretion rate are
correlated with the configuration of the magnetosphere.
Low values of $B_z$ at the position $(\varpi,z)=(0.6,0.4)$, correspond to
a configuration were closed magnetospheric lines penetrate the inner disc edge,
thus connecting the star to the disc
(as in the left hand panel of Fig.~\protect\ref{FRun_mag_strong-p-98}).
In this case, disc matter is loaded onto magnetospheric field lines
and flows along them to accrete onto the star.
As a consequence, the accretion rate is highest during these phases,
typically up to between
$10^{-8}M_\odot\yr^{-1}$ and $2.5\times10^{-8}M_\odot\yr^{-1}$.

High values of $B_z$ at the position $(\varpi,z)=(0.6,0.4)$, correspond to
a configuration when the outer field lines of the magnetosphere have opened up
into disconnected open stellar and disc field lines,
thus disconnecting the star from the disc
(as in the right hand panel of Fig.~\protect\ref{FRun_mag_strong-p-98}).
In this case, matter is lost directly into the outflow and
there is no net accretion of disc matter.

\section{Conclusions}

Simulations of disc dynamos as well as of star-disc coupling have shown
several surprises that were not originally expected.
On the other hand, some of these results are still only tentative
and one cannot really be sure to what extent they are artifacts of the
model setup or other technical limitations.
The fact that $\alpha_{\rm dyn}$ is negative in the upper disc plane
is one such result that has so far only been obtained in the local
shearing box simulations.
It would be good to check this result independently using global models.
Most of the recent global simulations tend to focus on the magnetic
field structures and their effects on the gas rather than on studying
the magnetic field {\it generation}, for example by identifying
parameterizations that could be used for mean field modeling.

Regarding star--disc coupling there are questions concerning the
strength of numerical viscosity and other technical aspects that
make it difficult to be sure about the significance of the result.
Here, however, the basic processes involved are now well understood,
in particular the inflation of field lines that are being
differentially sheared are they permeate the disc.
As argued by Matt \& Pudritz (2004, 2005), this makes the disc-locking
of the star rather inefficient.
This motivates the search for other mechanisms facilitating stellar braking.
A plausible candidate is braking by a stellar wind, which was also found
by von Rekowski \& Brandenburg (2006).
Extending these types of calculations to three dimensions and to
higher resolution remain important future tasks
(see, e.g., von Rekowski \& Piskunov 2006 for initial studies).
 
\begin{acknowledgements}
Use of the supercomputer SGI 3800 in Link\"oping and of the PPARC supported
supercomputers in St Andrews and Leicester is acknowledged. This research was
conducted using the resources of High Performance Computing Center North
(HPC2N).
The Danish Center for Scientific Computing is acknowledged
for granting time on the Linux cluster in Odense (Horseshoe).
\end{acknowledgements}

\bibliographystyle{aa}

\end{document}